\documentclass[twocolumn,showpacs,preprintnumbers,amsmath,amssymb]{revtex4}

\usepackage{graphicx}
\usepackage{psfig}
\usepackage{dcolumn}
\usepackage{bm}

\begin{document}

\title{
Instabilities of switching processes in
synthetic antiferromagnets
}

\author{A.N.\ Bogdanov}
\altaffiliation[Permanent address: ]%
{Donetsk Institute for Physics and Technology,
340114 Donetsk, Ukraine
}
\email{bogdanov@ifw-dresden.de}
\author{U.K. R\"o\ss ler}
\thanks
{Corresponding author 
}
\email{u.roessler@ifw-dresden.de}

\affiliation{
IFW Dresden,
Postfach 270116
D--01171 Dresden, Germany
}%

\date{\today}

\begin{abstract}
{
It is shown that magnetic states 
and field-driven reorientation transitions 
in synthetic antiferromagnets
crucially depend on contributions of higher-order anisotropies.
A phenomenological macrospin model is derived to describe 
the magnetic states of two antiferromagnetically coupled magnetic 
thin film elements. The calculated phase diagrams show 
that magnetic states with out-of-plane magnetization, 
symmetric \textit{escaped spin-flop} phases,
exist in a broad range of the applied magnetic field.
Due to the formation of such states and concomitant
multidomain patterns, the switching processes in toggle
magnetic random access memory devices (MRAM) can radically deviate 
from predictions within oversimplified models.
}
\end{abstract}

%

         
\maketitle

%
Recently, a controlled switching 
of synthetic antiferromagnetic thin-film elements 
has received high interest as active part of so-called 
\textit{toggle} magnetic random access memory systems 
\cite{Maffitt06}.
In these devices, an antiferromagnetically coupled
two-layer system undergoes a field-driven reorientation.
%
%
Various intrinsic and induced magnetic couplings 
in nanomagnetic layer systems lead to complex magnetic 
behaviour characterized by an unusually large number of 
magnetic states and transitions between 
them (see, e.g., \cite{PRB04}).
Therefore, even simplified 
phenomenological single-domain (or macro-spin) models 
display a great many magnetic states 
and complex magnetization processes \cite{Worledge04,Wang05}.
%
%
However, a general analysis can be given 
in terms of the \textit{topologically} different 
types of magnetic phase diagrams
to identify the \textit{stability} of 
magnetic states with respect to different 
perturbing interactions and 
the occurrence of \textit{metastable} states.
In this Letter, we show that higher-order
anisotropies give rise to additional magnetic 
states with {\em out-of-plane magnetization} 
in synthetic antiferromagnets.
Qualitatively, this effect is easily understood.
The field-driven reorientation transition 
in a synthetic two-layer antiferromagnet 
with a common uniaxial anisotropy 
is akin to the spin-flop transition 
in a classical two-sublattice antiferromagnet. 
At the spin-flop field, the uniaxial anisotropy and
the Zeeman energy balance each other. 
Hence, near the spin-flop field, the magnetic state 
is subject to weaker higher-order anisotropies.
This may cause magnetic instabilities. 
Here, we analyse the instability 
mechanism for an elementary macrospin model 
with competing uniaxial and cubic anisotropy. 
This  model is relevant for many magnetic layer systems 
with an intrinsic cubic magnetocrystalline anisotropy 
and an in-plane easy axis, e.g. induced by substrate-film interactions.
However, such instabilities of the spin-flop phase
in an antiferromagnet are 
a general phenomenon related 
to higher-order magnetic anisotropies.
E.g., a spin-flop phase transverse to 
the easy axis and an obliquely applied field 
can also be stabilized by  (intrinsic or induced) 
fourth order uniaxial anisotropy,
as has been observed and 
analysed quite a long time 
ago for bulk Cr \cite{Barak81}.
Similar {\em escaped spin-flop} states arise 
due to a secondary easy axis.
In layered synthetic antiferromagnets, 
such a biaxial anisotropy can be induced by 
perpendicular surface anisotropy
of the single ferromagnetic layers
\cite{Neel54,Bogdanov02,JMMM05}.

We consider spatially homogeneous states in
two physically \textit{identical} ferromagnetic
layers with a circular cross section (or infinite
in the basal plane) and an exchange-coupling
through an interlayer.
The phenomenological energy for this two-layer system
can be written as \cite{PRB04}
\begin{eqnarray}
W &=& J\, \mathbf{m}_1 \cdot \mathbf{m}_2 
- J\,\mathbf{h} \cdot (\mathbf{m}_1 + \mathbf{m}_2)
+\sum_{i=1}^2W_a^{(i)}\,,
\label{energy2}
\end{eqnarray}
where the unity macrospins
$\mathbf{m}_i$ = $\left\langle \mathbf{M}_i \right\rangle$/$M_0$ 
point along the average of the magnetization 
distribution $\mathbf{M}_i$, which is 
given by $\left\langle \mathbf{M}_i \right\rangle= 
\int_{\Omega_i} \mathbf{M}_i\; d^3r_i$,
$M_0 =|\left\langle \mathbf{M}_i\right\rangle|$,
for the $i$th layer with volume $\Omega_i$. 
The effective interlayer exchange is antiferromagnetic,
$J = M_0^2\;\int_{\Omega_1}\int_{\Omega_2} j (\mathbf{r}_1,\mathbf{r}_2) 
 d^3r_1\,d^3r_2 > 0$.
In the Zeeman term, 
the reduced (internal) field $\mathbf{h} = \mathbf{H}/H_e$ 
with $H_e=J/M_0$ comprises applied and stray fields.
The magnetic anisotropies should 
also be considered as volume averages 
\(W_a{(i)}=\int_{\Omega_i}\,w_i\; d^3r_i\).
The form of energy (\ref{energy2}) coincides
with the (mean-field) energy of 
a bulk two-sublattice antiferromagnet with 
``sublattice magnetizations'' $\mathbf{m}_i$
(see review and bibliography in \cite{PRB02,UFN88}).
The magnetic anisotropy of our model
considers a uniaxial ($K$) and a cubic ($K_c$)
contribution
$W_a^{(i)} = -K(\mathbf{m}_i \cdot \mathbf{n})^2/2
 -K_c [(\mathbf{m}_i \cdot \mathbf{a}_i)^4
 +(\mathbf{m}_i \cdot \mathbf{a}_2)^4
 +(\mathbf{m}_i \cdot \mathbf{a}_3)^4]/4$.
\begin{figure}[tbh]
\includegraphics[width=7.0cm]{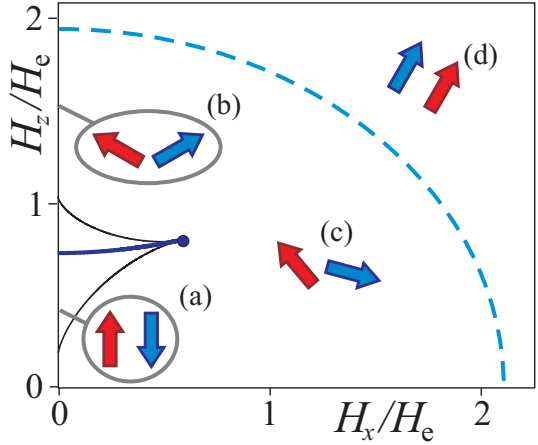}
%
\caption{
\label{PhD}
The magnetic phase diagram for small cubic
anisotropies $K_c < K $ holds only 
planar stable phases.
Antiferromagnetic (a) and spin-flop (b) states
exist in a magnetic field $H_z$ 
along the easy axis $\mathbf{n}$. 
The first-order (\textit{spin-flop}) transition
occurs in the critical field
$h_{SF}$.
In oblique magnetic fields,  $H_x \neq 0$,
the transition occurs between canted (c) phases
that are distorted antiferromagnetic
and spin-flop states.
Stability limits of metastable states are
given by thin lines.
Dashed line gives the continuous 
transition into the spin-flip
(saturated ferromagnetic) phase. 
}
\end{figure}

By a standard procedure we introduce
the vector of the total magnetization
$ \mathbf{M} = \mathbf{m}_1 + \mathbf{m}_2$
and the staggered vector
$ \mathbf{L} = \mathbf{m}_1 - \mathbf{m}_2$.
Because $|\mathbf{m}_i|$ =1,
the following constraints hold: 
$ \mathbf{M} \cdot \mathbf{L}$ =0
and $\mathbf{M}^2 + \mathbf{L}^2 $ =1 .
For weak anisotropy $J \gg K, K_c$ 
the magnetic energy Eq.~(\ref{energy2})
can be simplified as
\(W = J\, (2\mathbf{M}^2-1) - 2J\,\mathbf{h} \cdot \mathbf{M}
 -K [ ( 1- \mathbf{M}^2) (\mathbf{l} \cdot \mathbf{n})^2 
 +(\mathbf{M} \cdot \mathbf{n})^2 ]
-K_c[(\mathbf{l} \cdot \mathbf{a}_1)^4
+(\mathbf{l} \cdot \mathbf{a}_2)^4
+(\mathbf{l} \cdot \mathbf{a}_3)^4]/2\),
where $\mathbf{l} = \mathbf{L}/|\mathbf{L}|$ 
is the unity vector along the staggered vector.
An independent minimization with respect to $\mathbf{M}$ 
yields
$\mathbf{M} =  \left[\mathbf{h} - 
(\mathbf{h} \cdot \mathbf{l})\mathbf{l} \right]/2$
(for details, see similar calculations in bulk
systems, e.g. \cite{PRB02}).
After substituting this expression for $\mathbf{M}$ 
and omitting constant terms,
one obtains the magnetic energy 
as function of the unity vector
\(\mathbf{l} \equiv%
(\sin \theta \cos \phi, \sin \theta \sin \phi, \cos \theta)\) 
as
\begin{eqnarray}
\widetilde{W}  &=& (J-(K/4)\,\cos(2\, \theta)\, (\mathbf{h}\cdot \mathbf{l})^2
 -K \cos^2 \theta
 \nonumber\\
& & -K_c \left( 1 + \cos^2 2\theta - \sin^4 \theta \sin^2 2 \phi
\right)/4 ,
\label{energy4}
\end{eqnarray}
for a configuration with the uniaxial direction $\mathbf{n}$ 
and one of the cubic axis $\mathbf{a}_i$ along the $Z$ axis,
and for magnetic fields
restricted to the $XOZ$ plane, 
\(\mathbf{h} \equiv (h \sin \psi, 0, h \cos \psi)\).
%
%
Depending on the ratio between uniaxial and cubic anisotropy,
there are three topologically
different magnetic phase diagrams

I. For $ K > K_c$ only planar phases with $\phi =0$ 
are stable, i.e., the magnetization vectors are in $XOZ$ plane.
In this case  energy~(\ref{energy4})
can be reduced to the following functional
\begin{eqnarray}
\Phi (\theta) = -\frac{\widetilde{K}}{2}\cos^2 2 \theta 
-\widetilde{H}_z \cos 2 \theta 
-\widetilde{H}_x \sin 2 \theta\,,
\label{Phi}
\end{eqnarray}
where $\widetilde{K} =K h^2 +2K_c $,
$\widetilde{H}_z=-J \left(h^2 \cos 2 \psi - h_{SF}^2 \right)$,
$\widetilde{H}_x=Jh^2 \sin 2 \psi$, and
$h_{SF} = \sqrt{2K/J}$ is the reduced \textit{spin flop} field.

The energy (\ref{Phi}) is akin to 
the free energy of a bulk uniaxial ferromagnet \cite{Hubert98} 
and of antiferromagnets \cite{UFN88}.
The corresponding magnetic phase diagram 
is shown in Fig. \ref{PhD}.
At zero field and in a magnetic field along
the $Z$ axis,  $h < h_{SF}$, the antiferromagnetic
phase (Fig. \ref{PhD} (a)) has the lowest energy
while the \textit{spin-flop} phase ($\mathbf{l} || X$ axis,
Fig. \ref{PhD} (b)), has lowest energy for 
$h_{SF} < h < h_1= 2 - K_c/J -K/J$.
At the field $h_{SF}$ a proper \textit{spin-flop} takes place.
This is the first-order transition
between the collinear and the flopped states.
As the field approaches $h_1$ 
the spin-flop phase continuously transforms into the
saturated \textit{spin-flip} phase, Fig. \ref{PhD} (c).
In magnetic fields with components transverse to
the easy direction, 
a canted phase is stable, Fig. \ref{PhD} (d).
The region of the metastable states 
near the spin-flop transition is 
circumscribed by a modified astroid 
\((\widetilde{H}_x)^{2/3}+(\widetilde{H}_z)^{2/3}
=(\widetilde{K})^{2/3}\).
The hyperbola $h^2 \cos 2 \psi = h_{SF}^2$
gives the line of the first-order transition
between the distorted antiferromagnetic phase,
$\theta_1 = \arcsin(\tan 2 \psi /\tan 2 \psi_c)/2$,
and the spin-flop phase, $\pi/2 -\theta$,
where the maximum angle $\psi_c = \arctan(K/J + K_c/K)/2$
designates the end point of the first-order transition line,
Fig. \ref{PhD}.
\begin{figure}[htb]
\includegraphics[width=7.5cm]{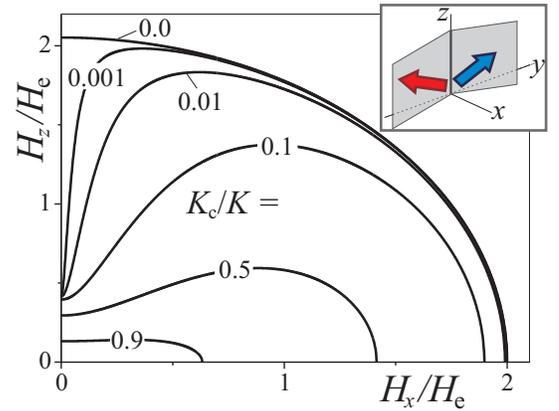}
%
\caption{
\label{lability}
Lability lines of the escaped spin-flop phase 
for $K/J =0.1$ and different ratios between cubic and 
uniaxial anisotropy.
}
\end{figure}

In a purely uniaxial system, $(K_c =0)$,
the spin-flop state has an infinite
degeneracy with respect to rotation 
of the vector $\mathbf{l}$ around the $Z$ axis. 
Application of a magnetic field 
with a component transverse to the easy axis 
lifts this degeneracy.
Then only a canted phase exists with
the magnetization  vectors in the plane
spanned by the easy direction and $\mathbf{H}$ 
(the ($XOZ$) plane in Fig. \ref{PhD}).
\begin{figure}[bth]
\includegraphics[width=7.5cm]{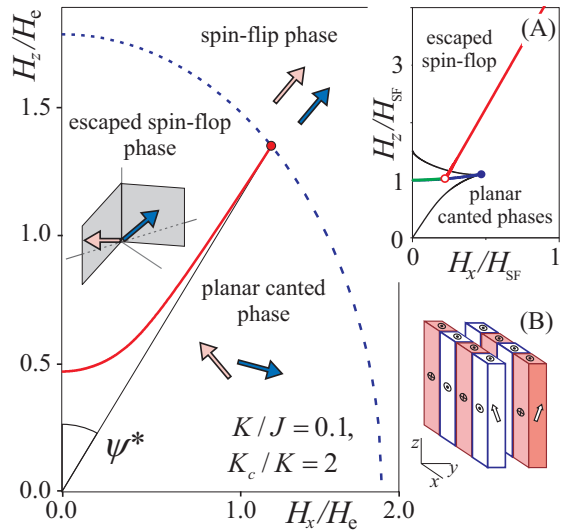}
%
\caption{
\label{PhD2}
The magnetic phase diagram for
$K < 3\, K_c/4$ 
contains a first order transition,
thick (red) line,
between the canted and escaped spin-flop phase. 
Both phases continuously
transform into the saturated state
at critical lines (dashed blue).
The first-order and two second-order lines meet in
a bicritical point.
Inset (A):
the phase diagram for case $K_c  > K > 3\, K_c/4$,
is intermediate between the simple case planar $K > K_c$ and the 
main figure.
%
Inset (B): sketch of a multidomain 
pattern in the escaped spin-flop phase.
}
\end{figure}

A remarkably different situation occurs for nonzero $K_c$.
The cubic anisotropy reduces the degeneracy
of the spin-flop phase to two preferable directions
in the $XOY$-plane, $\mathbf{a}_x$ and $\mathbf{a}_y$ 
in our model.
For finite $K_c$, a small enough component of 
the magnetic field along the $X$ axis 
can not destroy the local stability of 
the  spin-flop configurations with $\mathbf{l} || Y$.
This implies that cubic anisotropy can
stabilize spin-flop states with magnetization 
components out of the plane,
which we call a symmetric \textit{escaped spin-flop } state.

In the limit of weak anisotropies
the escaped spin-flop configuration 
can be written as
\begin{eqnarray}
 \mathbf{m}_{1}(\mathbf{m}_{2}) = 
 (h \sin(\psi)/2, \pm (\mp) \sqrt{4- h^2}/2, h \cos (\psi)).
\label{escapedS}
\end{eqnarray}
Note that the escaped spin-flop configuration (\ref{escapedS})
is degenerate with respect to the sign of the $y$-components.

The lability line for the escaped spin-flop phase can be
readily derived from Eq. (\ref{energy4})
\begin{eqnarray}
h_z^2= \frac{2J \left[4(1-(K_c/K))-h_x^2 \right]\left[h_x^2+ 2 K_c/J \right]}
{h_x^2(2J-K) +4 (K_c/K) (2J+3K))}\,.
\label{labilityESF}
\end{eqnarray}
For weak cubic anisotropy, $K_c/K \ll 1$,
the stability region of the escaped phase is
restricted to a narrow strip near the $H_z$ axis
and the vicinity of the spin-flip field
(Fig. \ref{lability}).
This region gradually spreads with increasing $K_c/K$,
and for $K_c/K >1$ it  covers the whole phase 
plane below the spin-flip line.

II. $ K_c > K > 3 K_c/4$
For $ K_c > K$ the escaped phase has
lowest magnetic energy and, therefore, it becomes
stable 
in certain regions of the phase diagram (Fig. \ref{PhD2}).
The first-order transition line between
the escaped and canted phase,  red lines in Fig. \ref{PhD2},
is given by a set of parametric equations
\begin{eqnarray}
(h_x,h_z) = \sqrt{\frac{2K}{J\,(1- \nu \sin^2 \theta)}} 
\left(\nu \sin \theta \cos^3 \theta,
     1- \nu \sin^2 \theta\right)
\label{ESF}
\end{eqnarray}
with $\nu=K_c/K$ and the solution $\theta$ 
for the canted phase at the transition line.
For $ 1 < K_c/K < 4/3$,
the transition line between the canted
and the escaped states consists of
a low field line and a high field line 
(green and red in Fig. \ref{PhD2} (A)).
These lines meet each other and the
first-order line between two differently canted
phases (blue) in a  \textit{triple} point 
($h_x^{(tr)} = \sqrt{2(K_c-K)/J}$,
$h_z^{(tr)} = \sqrt{2K_c/J}$).

III. $ K < 3 K_c/4$.
For $K_c = 4 K/3$,
the triple point merges with
the end point of the transition
between the canted phases.
The topology of the
phase diagram changes
(Fig. \ref{PhD2})
and now includes only one first-order
transition line between the escaped
and the canted planar phases.
This first-order  transition line (Eq.(\ref{ESF}))
ends at a \textit{bicritical} point
%
%
(\( h_x^* = 2 \sqrt{(K_c -K)/K_c}, \quad 
h_z^* = 2 /\sqrt{K/K_c}\),)
%
with $\psi^* = \arctan \sqrt{(K_c-K)/K}$,
where two continuous spin-flip lines meet.

The degeneracy of the escaped phases (\ref{escapedS}) 
with respect to the out-of-plane components of the 
magnetization may lead to the formation of regular 
multidomain states (Fig. \ref{PhD2} (B)).
The demagnetization in these domain structures
effectively counteracts the antiferromagnetic coupling.

The results for the model with 
a competition of uniaxial and cubic anisotropies  
exemplify the occurrence of escaped magnetic configurations 
in antiferromagnetic layer systems.
The escaped phase is metastable 
in wide regions of the field
even for weaker cubic anisotropies.
Therefore, switching processes 
in toggle mode MRAM devices may fail because 
the field-driven system can escape into 
metastable states with 
perpendicular magnetization components 
and the ensuing strong demagnetization effects.
%

{\em Acknowledgments.}
Work supported by DFG (RO 2238/6).
A.N.B.\ thanks H.\ Eschrig for support and
hospitality at IFW Dresden. 
%


\end{document}